\documentclass{webofc}
\usepackage[varg]{txfonts}   
\usepackage{physics}
\usepackage[capitalise]{cleveref}
\usepackage{xspace}

\begin{document}
\title{Progress in the Partial-Wave Analysis Methods at \mbox{COMPASS}}

\author{
\firstname{Florian Markus} \lastname{Kaspar}\inst{1,2}\fnsep\thanks{\email{florian.kaspar@tum.de}} \and
\firstname{Julien} \lastname{Beckers}\inst{1}\fnsep\thanks{\email{julien.beckers@tum.de}} \and
\firstname{Jakob} \lastname{Knollm\"uller}\inst{1,2}\fnsep
\thanks{\email{jakob.knollmueller@tum.de}}
\\ \emph{for the COMPASS Collaboration}
}

\institute{Technische Universit\"at M\"unchen, Physik Department, James-Franck-Straße 1, 85748 Garching bei M\"unchen
\and
           Excellence Cluster Origins, Boltzmannstraße 2, 85748 Garching bei M\"unchen
          }

\abstract{%
     We study the excitation spectrum of light and strange mesons in diffractive scattering. We identify different hadron resonances through partial wave analysis, which inherently relies on analysis models.
     Besides statistical uncertainties, the model dependence of the analysis introduces dominant systematic uncertainties.
     We discuss several of their sources for the $\pi^-\pi^-\pi^+$ and $K^0_S K^-$ final states and present methods to reduce them.
     We have developed a new approach exploiting a-priori knowledge of signal continuity over adjacent final-state-mass bins to stably fit a large pool of partial-waves to our data, allowing a clean identification of very small signals in our large data sets.
     For two-body final states of scalar particles, such as $K^0_S K^-$, mathematical ambiguities in the partial-wave decomposition lead to the same intensity distribution for different combinations of amplitude values.
     We will discuss these ambiguities and present solutions to resolve or at least reduce the number of possible solutions.
     Resolving these issues will allow for a complementary analysis of the $a_J$-like resonance sector in these two final states.
}
\maketitle
\section{Introduction} \label{sec:intro}
    The study of light-meson resonances is part of the physics program of the COMPASS experiment. They are produced as short-lived intermediate states in diffractive reactions between a hadron beam and the target. The spectrometer measures the full kinematics of the decay particles, allowing us to infer the light-meson spectrum by fitting the collected data.
    A thorough statistical analysis is required to reduce systematic uncertainties originating from the modelling procedure. It is therefore crucial to conduct detailed studies of the methodology.
    We present our recent advances in partial-wave analysis (PWA) methods at COMPASS using simulated data for two reactions, $\pi^- + p \to \pi^-\pi^-\pi^+ + p$ and $\pi^- + p \to K^0_S K^- + p$.
    In this paper, we provide more details on the methods developed for the first reaction. A more in-depth review of the second one will be given in \cite{hadron_proceedings}.

\section{Partial-Wave Analysis} \label{sec:PWA_methods}
    We model the observed data to extract information about the possible resonances.
    In the following, we summarize the method detailed in \cite{compass_review}.
    We assume pion diffractive dissociation as the dominant production mechanism for the intermediate states and describe them as a coherent superposition of contributions with distinct quantum numbers, so-called \emph{partial waves}.
    The total intensity is the magnitude squared of the sum of the complex-valued amplitudes and an incoherent background $\left| \mathcal{T}_\text{flat}(m_X) \right|^2$.
    \begin{equation} \label{eq:intensity_decomp}
        \mathcal{I}(m_X;\tau_n) = \left| \sum_{a} \, \mathcal{T}_a(m_X) \, \psi_a(m_X,\tau_n) \, \right|^2 + \left| \mathcal{T}_\text{flat}(m_X) \right|^2
    \end{equation}
    For each partial wave $a$, we can model the decay amplitude $\psi_a$
    describing the final-state-particle distribution in the phase-space variables $\tau_n$.
    The corresponding production amplitudes $\mathcal{T}_a$, together, contain the information about the intermediate states $X^-$ and are inferred from data.

    From \cref{eq:intensity_decomp} and the data, we obtain a likelihood which we can maximize with respect to the production amplitudes $\mathcal{T}_a$, usually in kinematic bins of the invariant mass $m_X$.

    The amplitudes obtained in this quasi-model-independent way then serve as input for a second fit, in which we extract the resonance parameters by modelling the amplitudes' mass dependence.
    
     \begin{figure*}
     \centering
     \sidecaption
     \includegraphics[width=0.3\linewidth,clip]{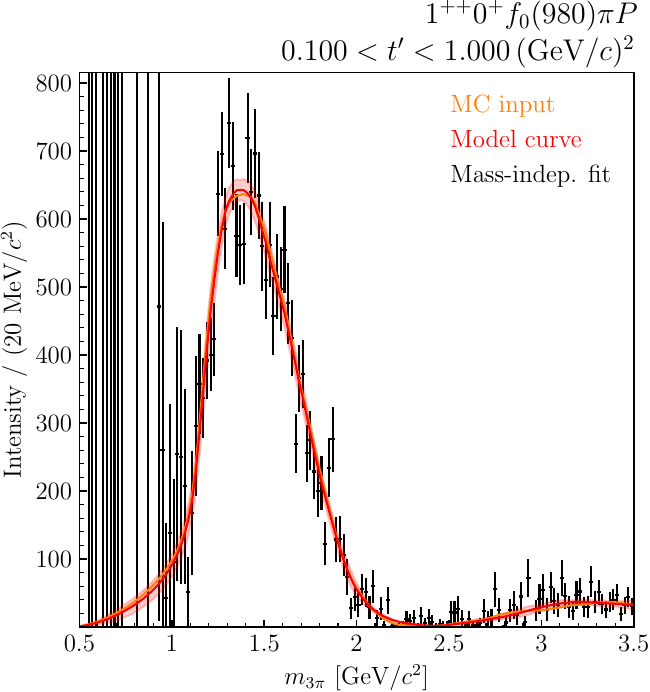}
     \includegraphics[width=0.3\linewidth,clip]{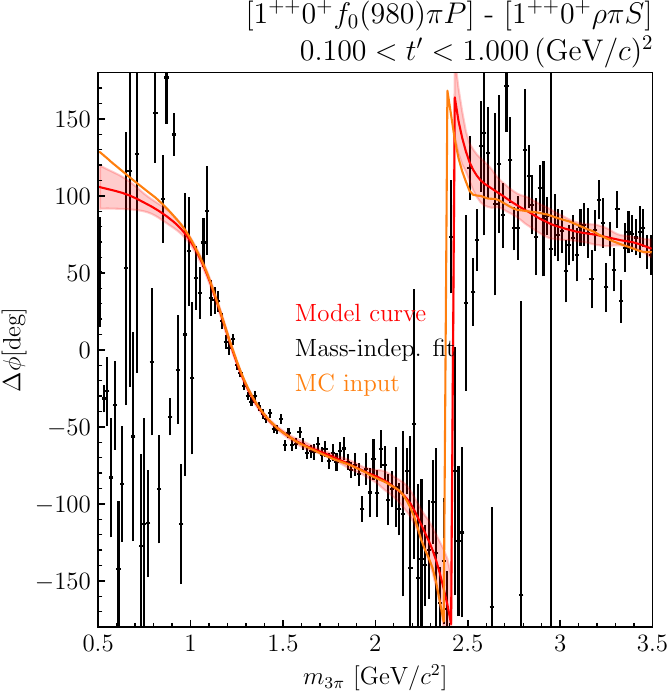}
     \caption{Comparison of a binned partial-wave analysis (black points) of simulated
     data with a resonance-model fit using the new method (red curve). The input model is in orange. \emph{Left:} comparison of the intensity. \emph{Right:} comparison of the relative phase.}
     \label{fig-2}       
     \end{figure*}

     \begin{figure*}
     \centering
     \sidecaption
     \includegraphics[width=0.3\linewidth,clip]{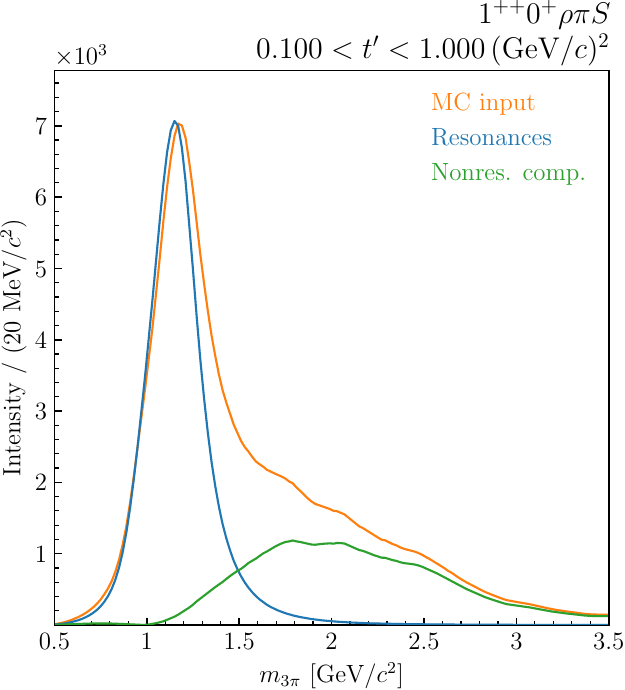}
     \includegraphics[width=0.3\linewidth,clip]{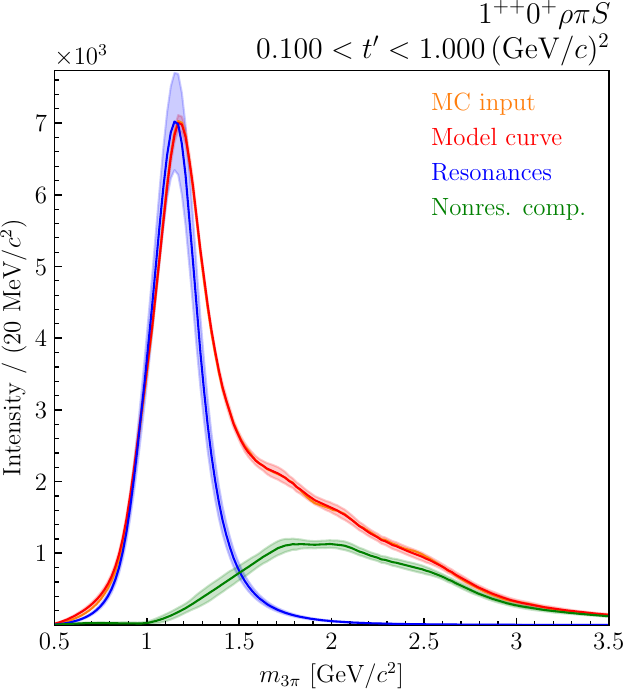}
     \caption{\emph{Left:} Input resonant (blue) and non-resonant (green) components that coherently add up to the total input model (orange). \emph{Right:} Results of the fit procedure. Recovered resonant (blue) and non-resonant (green) components. The resulting total signal lies perfectly on top of the input signal (orange). The  bands mark $\pm$ one standard deviation. (additional example see \cite{hadron_proceedings})}
     \label{fig-3}       
     \end{figure*}

\section{Continuity for Partial-Wave Analysis} \label{sec:continuity}
     The number of partial waves, and thereby the number of fit parameters, can be in the order of hundreds.
     Determining $\mathcal{T}_a(m_X)$ in separate bins may cause large statistical fluctuations and instabilities due to local maxima of the likelihood.
     Introducing additional information regularizes the fit and may thereby improve it.
     We expect the production amplitudes to be continuous over the kinematic variables such as $m_X$. Additionally, the reaction kinematics suppress the amplitudes at threshold and at high values of $m_X$.

     We include this prior knowledge into our model of the production amplitudes with the help of Information Field Theory (IFT) \cite{ift}.
     Instead of inferring $ \mathcal{T}_a $ in individual bins of $m_X$ independently, we construct a continuous model based on the correlated field model described in \cite{corr_fields} to determine a smooth, non-parametric solution in all bins simultaneously.
     By adjusting the model parameters, we can choose, for example, how smooth the amplitude model should be.
     Since IFT is defined in a Bayesian setting, all model parameters have prior distributions that represent our degree of model uncertainty.
     We can draw samples from these prior distributions, set them as parameters of the model and thereby obtain an overall sample of the model.
     With these \emph{prior predictive checks}, we can decide if our choice of priors on the model parameters leads to sensible amplitude models.

     Additionally, we can extend our new model with explicit parameterizations for resonances. In this case, the overall amplitude is a coherent sum of the correlated fields and the parameterizations (e.g. a Breit-Wigner). The correlated fields then take the role of an effective background model. We also set priors for the parameters of the resonance models.

     We implemented this new analysis method using the NIFTy framework for Numerical Information Field Theory \cite{nifty}.
     We performed in-depth tests of the new method using input-output studies. 
     For this, we first draw a random sample from our priors and obtain a prior sample for the amplitudes, which is the input of the study. According to this input we simulate events for the three-pion final state in bins of $m_X$.
     We then tried to recover the known input using both the usual method of independent bins and the new continuous PWA method. \Cref{fig-2} shows a comparison of the two different fitting methods. We observe large statistical fluctuations for the method of independent bins, while the fit with a continuous model recovers the input very well in both intensity and phase. Furthermore, \Cref{fig-3} shows that not only are we able to recover the total input, but that the new method is also able to accurately separate the resonant and non-resonant components.

     We conclude that the additional information incorporated into this new PWA method reduces the statistical uncertainty of the fit and even allows us to extract the resonance parameters directly in a single step, using a flexible and non-parametric background model.

\section{Ambiguities in Two-Body Final States} \label{sec:amb}

\begin{figure*}
        \centering
        \sidecaption
        \includegraphics[width=0.35\linewidth,clip]{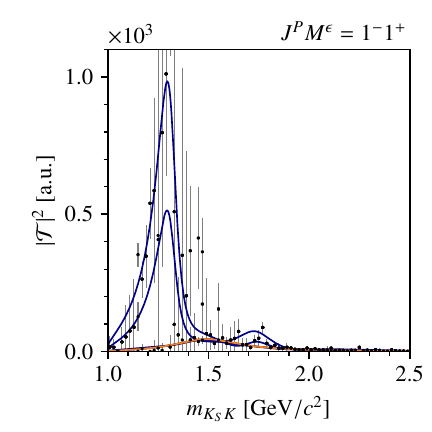}
        \includegraphics[width=0.35\linewidth,clip]{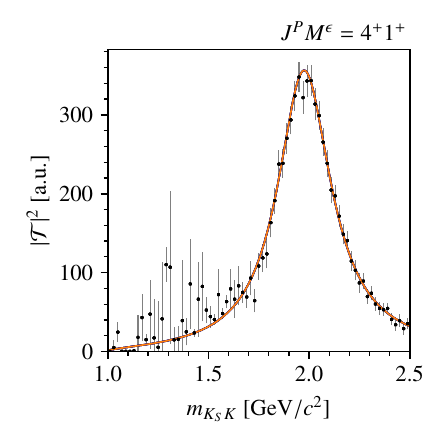}
        \caption{Partial-wave intensities. The model and calculated solutions are drawn as orange and blue curves respectively. The intensities obtained in the PWA fit are shown as black dots with grey error bars. (this and additional example can be found in \cite{hadron_proceedings})}
        \label{fig:amb_study_PWA}
    \end{figure*}

    For two-spinless-particle final states such as $K_S^0K^-$, the decomposition in \cref{eq:intensity_decomp} is not unique in each $m_X$ bin (see e.g. \cite{chung_ambiguities}).
    By expressing the decay amplitudes,\footnote{We use the reflectivity basis \cite{chung_spinDensityMatrix} and thus neglect contributions with $M \neq 1$ because of the dominance of Pomeron exchange at COMPASS energies \cite{compass_review} and the suppression of higher $M$ contributions. We drop indices $M^\varepsilon=1^+$.} which are simply spherical harmonics $Y_{J}^1$, in terms of $u\equiv \tan{(\theta/2)}$, the sum in \cref{eq:intensity_decomp} can be written as a polynomial of degree $J$ in $u^2$.
    Only its absolute value enters \cref{eq:intensity_decomp}, and the intensity hence remains invariant under complex conjugation of any combination of the polynomial's roots (called Barrelet Zeros \cite{barrelet}).
    As their values depend non-linearly on the amplitude values $ \{ \mathcal{T}_a \} $, this leads to different $ \{ \mathcal{T}_a \} $ which all yield the same intensity distribution in phase space.
    We can compute all ambiguous amplitude values from a starting set $ \{ \mathcal{T}_a \} $ by obtaining the roots numerically.
    We have studied the ambiguities using a continuous model in $m_X$ for four partial-wave amplitudes. 
    We first compute the exact distributions of the ambiguous amplitudes, shown as continuous curves in \cref{fig:amb_study_PWA}. We note that they are also continuous. In a second step, we simulate pseudodata according to the amplitude model and perform a PWA.\footnote{To find all solutions, we perform a large number of fitting attempts with random starting values.} The results are shown in \cref{fig:amb_study_PWA}, and we see that overall, the result of the fit is compatible with the calculated solutions, but the finite data reduces the number of solutions in some $m_X$ bins. We also observe that the highest-spin wave (in our case, $J^P=4^+$) is not affected by the ambiguities.
    
\section{Outlook} \label{sec:outlook}

    We are currently investigating the application of the continuity constraints to analyses of two-body final states. Initial promising results on simulation data suggest that this approach could separate the different ambiguous solutions. We are also actively applying the presented methods to COMPASS data to obtain new spectroscopic results with improved precision.

\section{Acknowledgements} \label{sec:acknowledgements}

     We would like to thank Philipp Frank and Torsten En{\ss}lin (both Max-Planck Institute for Astrophysics), as well as Stefan Wallner (Max-Planck Institute for Physics), who, together with the authors, made a first attempt to adapt NIFTy for partial-wave analyses. 
     We also thank the COMPASS Hadron Subgroup.\\
     Some of the results in this publication have been derived using the NIFTy package \cite{nifty}.\\
     Funded by the DFG under Germany's Excellence Strategy - EXC2094 - 390783311 and BMBF Verbundforschung 05P21WOCC1 COMPASS.

\end{document}